\documentclass[two column]{article}
\usepackage{graphicx}
\usepackage{subfigure}
\usepackage{amsfonts}
\usepackage{amsmath}
\newtheorem{proposition}{Proposition}

\begin{document}

\title{Gaussian entanglement witness and refined Werner-Wolf criterion for continuous variables}


\author{Xiao-yu Chen, Maoke Miao, Rui Yin, Jiantao Yuan\\
{\small {School of Information and Electrical Engineering, Zhejiang University City College, Hangzhou {\rm 310015}, China }}}


\date{}

\maketitle

\begin{abstract}
We use matched quantum entanglement witnesses to study the separable criteria of continuous variable states. The witness can be written as an identity operator minus a Gaussian operator. The optimization of the witness then is transformed to an eigenvalue problem of a Gaussian kernel integral equation. It follows a separable criterion not only for symmetric Gaussian quantum states, but also for non-Gaussian states prepared by photon adding to or/and subtracting from symmetric Gaussian states. Based on Fock space numeric calculation, we obtain an entanglement witness for more general two-mode states.  A necessary criterion of separability follows for two-mode states and it is shown to be necessary and sufficient for a two mode squeezed thermal state and the related two-mode non-Gaussian states. We also connect the witness based criterion with Werner-Wolf criterion and refine the Werner-Wolf criterion.
\end{abstract}

\section {Introduction}
Quantum entanglement is currently considered a central resource in quantum computation and quantum information processing\cite{HorodeckiR}. It is therefore crucial to be able to determine whether a quantum state is separable or entangled,  which is not an easy problem when it comes to mixed states. Many entanglement criteria have been proposed to detect entanglement\cite{HorodeckiR}\cite{Guhne}. In continuous variable (CV) systems, a necessary criterion for the separability of any two-mode state has been derived\cite{Duan}\cite{Simon}, which turns out to be a necessary and sufficient criterion for Gaussian states. Criteria have also been built for multi-mode bipartite Gaussian states \cite{Werner}\cite{Wang} and general CV states \cite{Shchukin}\cite{Chen07}\cite{ZhangDa}, respectively. Most of them \cite{Simon}\cite{Wang}\cite{Shchukin}\cite{Chen07}\cite{ZhangDa} are results of positive partial transpose (PPT) criterion \cite{Peres}\cite{Horodecki}. The fail of PPT criterion is the existence of the bounded entangled state. Bound entangled Gaussian states with two modes at both parties were proposed\cite{Werner}.  Other criteria for CV systems are, uncertainty principle criterion\cite{Duan}\cite{Agarwal}\cite{Hillery}\cite{Nha}, computable cross norm and realignment criterion (CCNR) \cite{Jiang14}, entropy criterion \cite{Walborn}, local operator criterion \cite{Zhang} and so on. We will add to the list of separable criteria another method based on integral equation of characteristic function.

Basically, separable criterion is an inequality (including matrix inequality) fulfilled by all separable states, violation of it implies entanglement. Thus a separable criterion can be converted to an entanglement witness \cite{Guhne}\cite{SperlingVogel}. An entanglement witness is a Hermitian operator which has non-negative means on all separable states and has a negative mean at least on one entangled state. The definition of the entanglement witness leads to an inequality on the witness operator. The optimization of the inequality fixes one of the parameters of the witness, it is called weakly optimal \cite{Guhne}. A full optimization \cite{Chen17} of all the parameters of a witness will lead to a matched entanglement witness, which can detect all the entanglement for enough parameters.

The criteria mentioned above usually are better in performance for Gaussian states than non-Gaussian states. However, non-Gaussian states have been proven to be not only indispensable resources for universal quantum computation \cite{Niset}\cite{Walschaers}, but also demonstrate superior
performance in many continuous variable quantum information protocols, such as quantum key distribution \cite{Hu}, quantum metrology\cite{Strobel}\cite{Joo}, and quantum imaging \cite{Liu}. The precise detection of the entanglement of non-Gaussian states are crucial for these applications. We will provide necessary and sufficient criterion of separability for non-Gaussian states prepared by photon adding to or/and subtracting from Gaussian states.

The paper is organized as follows. In section \ref{sec2}, we introduce the elements of entanglement witness and characteristic function of continuous variable system. Section \ref{sec3} is devoted to the criterion from integral equation of characteristic function. In section \ref{sec4}, we present numeric method of optimizing the mean of Gaussian operator over product states for the purpose of validity of Gaussian witness. In section \ref{sec5}, we extend the criteria to non-Gaussian states prepared by photon adding to and/or subtracting from Gaussian states. In section \ref{sec6}, we relate criterion from Gaussian operator witness to Werner-Wolf criterion and refine Werner-Wolf criterion to a simplified form for applications. Discussion and conclusion are given in section \ref{sec7}.

\section{Preliminary}\label{sec2}
Assuming a witness with the form of $\hat{W}=\Lambda \hat{I}-\hat{M}$, where $\hat{I}$ is the identity operator. The validity of a witness requires that $Tr(\hat{\rho}_{s}\hat{W})=\Lambda-Tr(\hat{\rho}_{s}\hat{M})\geq0$ for all separable states $\hat{\rho}_{s}$. For any given Hermitian operator $\hat{M}$ (We will call it a detect operator), let $\Lambda=\max_{\hat{\rho}_{s}}Tr(\hat{\rho}_{s}\hat{M}),$  then $\hat{W}$ is a weakly optimal entanglement witness. If $Tr(\hat{\rho}\hat{W})=\Lambda-Tr(\hat{\rho}\hat{M})<0$ for a state $\hat{\rho}$, then $\hat{\rho}$ is an entangled state. Hence for a given detect operator $\hat{M}$, there is a necessary criterion of separability $Tr(\hat{\rho}\hat{M})\leq\Lambda.$
 For a given state $\hat{\rho}$, we define $\mathcal{L}=\min_{\hat{M}}\frac{\Lambda}{Tr(\hat{\rho} \hat{M})}$ subject to $\Lambda>0$ and $Tr(\hat{\rho} \hat{M})>0$. The refined separable criterion is
 \begin{equation}\label{we0a}
   \mathcal{L}\geq 1.
 \end{equation}
The extreme detect operator, $\hat{M^{*}}$, that minimizes $\frac{\Lambda}{Tr(\hat{\rho} \hat{M})}$, leads to a matched entanglement witness, $\hat{W^{*}}=\Lambda \hat{I}-\hat{M^{*}}$. Notice that (\ref{we0a}) is a necessary and sufficient criterion for separability. Since if $\mathcal{L}< 1$, then the state $\hat{\rho}$ is entangled. While if $\mathcal{L}\geq 1$, the state $\hat{\rho}$ should be separable. Otherwise, if there is an entangled state $\hat{\rho}$ such that $\mathcal{L}\geq 1$, it means that there is no witness can detect this entangled state. However, it has been proved that `for each entangled state $\hat{\rho}$, there exists an entanglement witness detecting it' \cite{Guhne}\cite{Horodecki}.

For CV systems, we use the notation $\hat{R}=(\hat{x}_{1},\hat{p}_{1},...,\hat{x}_{n},\hat{p}_{n})^T$ for an operator vector containing the position $\hat{x}_{j}$ and momentum $\hat{p}_{j}$ quadratures of all the modes $(j=1,...,n)$. The commutation relations read $[\hat{R}_{j},\hat{R}_{l}]=i\sigma_{jl}$, where the symplectic matrix takes the form $\sigma=\left(
                                                                               \begin{array}{cc}
                                                                                 0 & 1 \\
                                                                                 -1 & 0 \\
                                                                               \end{array}
                                                                             \right)
^{\oplus n}$. The covariance matrix (CM) $\gamma$ of a state $\hat{\rho}$ can be written in terms of $\gamma_{ij}=\langle\hat{R}_{i}\hat{R}_{j}+\hat{R}_{j}\hat{R}_{i}\rangle_{\hat{\rho}}-2\langle \hat{R}_{i}\rangle_{\hat{\rho}}\langle\hat{R}_{j}\rangle_{\hat{\rho}}$. A local displacement does not affect the entanglement, which implies that we only need to consider the case of $\langle \hat{R}\rangle_{\hat{\rho}}=0$. The characteristic function of a state is defined as $\chi(z)=:Tr[\hat{\rho}\exp(iz\hat{R})]$. A Gaussian state is completely characterized by its CM, that is, $\chi(z)=\exp{(-\frac{1}{4}z\gamma z^T)}$ when the first moment is set to zero.

For a bipartite or multipartite $n$-mode Gaussian state $\hat{\rho}_{G}$, we choose a zero-mean Gaussian operator $\hat{M}$ as the detect operator. The witness operator $\hat{W}$ is not a Gaussian operator. The CMs of $\hat{\rho}_{G}$ and $\hat{M}$ are assumed to be $\gamma$ and $\gamma_{M}$, respectively. Accordingly, we have $Tr(\hat{\rho}_{G}\hat{M})=2^{n}[\det(\gamma+\gamma_{M})]^{-\frac{1}{2}}$. In order to find $\Lambda$ to form a witness to detect Gaussian state entanglement, it suffices to consider Gaussian separable states $\hat{\rho}_{Gs}$, we get $\Lambda=\max_{\hat{\rho}_{Gs}} Tr(\hat{\rho}_{Gs}\hat{M})$. The difficulty of the optimization problem is that, although a separable Gaussian state can be decomposed as the probability summation of product Gaussian states, it can also be decomposed as the probability summation of non-Gaussian states. An example is the two-mode squeezed thermal state, which can be written as the probability summation of particle number product states acted on with two-mode squeezing. Therefore, we have to maximize $Tr(\hat{\rho}_{Gs}\hat{M})$ with respect to all the product states to obtain $\Lambda$, written by $\Lambda=\max_{\hat{\rho}_{A},\hat{\rho}_{B}}Tr[\hat{M}(\hat{\rho}_{A}\otimes\hat{\rho}_{B})]$, if the system is divided into A and B subsystems. It suffices to consider $\hat{\rho}_{A}$ and $\hat{\rho}_{B}$ to be pure states.

\section{Criterion from integral equation of characteristic function }\label{sec3}
The characteristic function of an operator $\mathcal{P}$ is defined as $\chi_{\mathcal{P}}(z)=Tr(\mathcal{P}e^{iz\hat{R}})$. Then $\Lambda=\max_{\chi_{A},\chi_{B}}\int \chi_{A}(z_{A})\overline{\chi}_{M}(z_{A},z_{B})\chi_{B}(z_{B})[dz_{A}][dz_{B}]$, where $\chi_{A}, \chi_{B}$ and $\chi_{M} $ are the characteristic functions of $\hat{\rho}_{A}, \hat{\rho}_{B}$ and $\hat{M}$, respectively. $\overline{\chi}_{M}$ is the conjugate of $\chi_{M}$, and  we have denoted $[dz]=\frac{1}{(2\pi)^n}\prod_{j=1}^n dz_{2j-1}dz_{2j}$. The characteristic function of a pure local subsystem state has the property of $\int|\chi_{j}(z_{j})|^2[dz_{j}]=Tr(\hat{\rho}_{j}^2)=1$, $j=A$ or $B$. Accordingly, the maximization problem is transformed into
\begin{eqnarray}
  &&\int\overline{\chi}_{M}(z_{A},z_{B})\chi_{B}(z_{B})[dz_{B}]=\lambda_{A}\overline{\chi}_{A}(z_{A}) \label{we0b}, \\
  &&\int\overline{\chi}_{M}(z_{A},z_{B})\chi_{A}(z_{A})[dz_{A}]=\lambda_{B}\overline{\chi}_{B}(z_{B}) \label{we0c},
\end{eqnarray}
where $\lambda_{A}$ and $\lambda_{B}$ are Lagrangian multipliers. Clearly, $\lambda_{A}=\lambda_{B}=\int \chi_{A}(z_{A})\overline{\chi}_{M}(z_{A},z_{B})\chi_{B}(z_{B})[dz_{A}][dz_{B}]:=\lambda$.
The combination of the two equations leads to an eigen-equation of symmetric Gaussian kernel, namely,
\begin{equation}\label{sp0}
   \int\chi_{2M}(z_{A},z'_{A})\chi_{A}(z'_{A})[dz'_{A}]=\lambda^2 \chi_{A}(z_{A})
 \end{equation}
 where $\chi_{2M}(z_{A},z'_{A})=\int\chi_{M}(z_{A},z_{B})\overline{\chi}_{M}(z'_{A},z_{B})[dz_{B}]=\frac{1}{\sqrt{\det(\gamma_{2})}}\exp{[-\frac{1}{4}(z_A,z'_{A})\gamma_{2M}(z_A,z'_{A})^T]}$ is a 2-fold symmetric Gaussian kernel with CM $\gamma_{2M}=\left(
                                                      \begin{array}{cc}
                                                        \zeta & \omega\\
                                                        \omega& \zeta \\
                                                      \end{array}
                                                    \right)
 $, with $\zeta=\gamma_{1}+\omega$,   $\omega=-\frac{1}{2}\gamma_{3}\gamma_{2}^{-1}\gamma_{3}^{T}$.
Where we define
\begin{equation}\label{sp1}
  \gamma_{M}=\left(
                                    \begin{array}{cc}
                                      \gamma_{1} & \gamma_{3} \\
                                      \gamma_{3}^T & \gamma_{2}\\
                                    \end{array}
                                  \right).
\end{equation}

The problem of maximizing the mean of a Gaussian detect operator over product states is reduced to find the maximal square root eigenvalue of the 2-fold kernel integral equation. For Gaussian detect operators with $\gamma_{1}$ and $\omega$ can be diagonalized by the same symplectic transformation, we will show in Appendix A that the maximal eigenvalue of the Gaussian kernel integral eigen-equation is achieved by a Gaussian eigenfunction, the non-Gaussian eigenfunctions are obtained by the functional derivatives of the Gaussian eigenfunction, their absolute eigenvalues are all less than the eigenvalue corresponding to the Gaussian eigenfunction. The idea of proof is to transform the Gaussian kernel into product of integral kernels, each of them has a series of eigenfunctions which are wave functions of an harmonic oscillator (see Appendix A for details). So the maximal mean of Gaussian detect operator over product state is achieved by Gaussian functions $\chi_{A}(z_{A})$ and $\chi_{B}(z_{B})$, and the extremal states, $\hat{\rho}_{A}^*$ and $\hat{\rho}_{B}^*$ are Gaussian states. Then, we have
$\Lambda=2^n[\det(\gamma_{A}\oplus\gamma_{B}+\gamma_{M})]^{-\frac{1}{2}}$,
where $\gamma_{A}$ and $\gamma_{B}$ are the CMs of local Gaussian states, $\hat{\rho}_{A}^{*}$ and $\hat{\rho}_{B}^{*}$, respectively. The CMs, $\gamma_{A}$ and $\gamma_{B}$, can be obtained directly from Eqs. (\ref{we0b}) and (\ref{we0c}), which are given by
\begin{eqnarray}
  && \gamma_{A}=\gamma_{1}-\gamma_{3}(\gamma_{2}+\gamma_{B})^{-1}\gamma_{3}^T,  \label{we0d}\\
  && \gamma_{B}=\gamma_{2}-\gamma_{3}^T(\gamma_{1}+\gamma_{A})^{-1}\gamma_{3}.  \label{we0e}
\end{eqnarray}

Besides the normalization conditions of the characteristic functions $\chi_{A}(z_{A})$ and $\chi_{B}(z_{B})$, there are further requirements of $\chi_{A}(0)=1$ and $\chi_{B}(0)=1$ since $\hat{\rho}_{A}^*$ and $\hat{\rho}_{B}^*$ are states. These conditions lead to $\det(\gamma_{A})=\det(\gamma_{B})=1$. The freedom in choosing $\gamma_{M}$ is quite limited by $\det(\gamma_{A})=\det(\gamma_{B})=1$ via equations (\ref{we0d}) and (\ref{we0e}).

Let $\Omega$  be the set of $\gamma_{M}$ such that (i) $\gamma_{1}$ and $\omega=0$ are simultaneously diagonalizable by some symplectic transformation (ii) the solution of (\ref{we0d}) and (\ref{we0e}) needs to guarantee $\det(\gamma_{A})=\det(\gamma_{B})=1$, namely, the local Gaussian states, $\hat{\rho}_{A}^{*}$ and $\hat{\rho}_{B}^{*}$ are pure states.
 We then have the following proposition from the definition of the witness.
\begin{proposition}\label{proposition1}
For a Gaussian state with CM $\gamma$, the necessary criterion of separability is
\begin{equation}\label{we1}
   \det(\gamma+\gamma_{M})\geq \det(\gamma_{A}\oplus\gamma_{B}+\gamma_{M}),
\end{equation}
for any given $\gamma_{M}$ with $\gamma_{M}+i\sigma\geq0$ and $\gamma_{M} \in \Omega$.
\end{proposition}
In order to keep $Tr(\rho \hat{M})>0$, we simply let $\hat{M}$ to be an unnormalized Gaussian density matrix, so that $\gamma_{M}+i\sigma\geq0$.
In the following, we will write $\gamma$ in block form for the partition $A$ versus $B$.
\begin{equation}\label{we2}
  \gamma=\left(
           \begin{array}{cccc}
             \mathcal{A} & \mathcal{C} \\
             \mathcal{C} & \mathcal{B} \\
           \end{array}
         \right),
\end{equation}

\subsection{Two-mode symmetric Gaussian state}
   The first example is a two-mode symmetric Gaussian state with CM given by (\ref{we2}) with $\mathcal{A}=\mathcal{B}=diag(a,a); \mathcal{C}=diag(c_{1},-c_{2})$ and $c_{1}>0, c_{2}>0$.  We choose $\gamma_{M}$ such that it has the similar structure of $\gamma$. It is
\begin{equation}\label{we2a0}
  \gamma_{M}=\left(
                                    \begin{array}{cccc}
                                      M_{1} & 0& M_{5}&0 \\
                                      0&M_{2} & 0& -M_{6} \\
                                      M_{5}&0& M_{3}&0\\
                                      0& -M_{6}&0& M_{4}
                                    \end{array}
                                  \right).
\end{equation}
The condition of $\gamma_{M}\in \Omega$ leads to
\begin{eqnarray}
  && M_{1}M_{2}=M_{3}M_{4}, \label{we2a}\\
  && (M_{1}M_{3}-M_{5}^2)(M_{2}M_{4}-M_{6}^2)=1. \label{we2b}
\end{eqnarray}
We further assume $M_{1}=M_{2}=M_{3}=M_{4}$. Proposition \ref{proposition1} leads to
\begin{eqnarray}
  &[a^2-c_{1}^2+2(aM_{1}-c_{1}M_{5})+M_{1}^2-M_{5}^2] \nonumber\\
  &\times[a^2-c_{2}^2+2(aM_{1}-c_{2}M_{6})+M_{1}^2-M_{6}^2]\nonumber \\
  &\geq 4[M_{1}+\sqrt{M_{1}^2-M_{5}^2}][M_{1}+\sqrt{M_{1}^2-M_{6}^2}].
\end{eqnarray}
   The right hand side of the inequality  is obtained by solving $\gamma_A$ and $\gamma_B$ using (\ref{we0d}) and (\ref{we0e}) explicitly. That is $\gamma_{A}=\gamma_{B}=diag(\sqrt{M_{1}^2-M_{5}^2},\sqrt{M_{1}^2-M_{6}^2})$. Thus $\rho_{A}^{*}$ and $\rho_{B}^{*}$ are squeezed vacuum states. In the limit of $M_{i}\rightarrow \infty$ while keeping $\sqrt{M_{1}^2-M_{5}^2}$ and $\sqrt{M_{1}^2-M_{6}^2}$ finite (this is possible by setting some finite squeezing parameter $r$ for $\rho_{A}^{*}$ and $\rho_{B}^{*}$), we have the separable condition
   \begin{equation}\label{we2a1}
    (a-c_{1})(a-c_{2})\geq 1,
   \end{equation}
    which is the necessary and sufficient separable criterion\cite{Duan}\cite{Simon} for a two mode symmetric Gaussian state.
Condition (\ref{we2b}) shows us that the Gaussian detect operator $\hat{M}$ is in fact an unnormalized density operator representing a pure quantum Gaussian state. Therefore, we use a two-mode infinite squeezing pure Gaussian state as a tool to detect entanglement.
\subsection{Werner-Wolf symmetric $2\times2$ state}
   We consider a symmetric $2\times2$ Gaussian state, which has a CM of the form (\ref{we2}) with $\mathcal{A}=\mathcal{B}=AI_{4}$ ($I_{n}$ is the $n\times n$ identity matrix) and
   \begin{equation}\label{we2a2}
     \mathcal{C}=\left(
                   \begin{array}{cccc}
                     E & 0 & 0 & 0 \\
                     0 & 0 & 0 & -F \\
                     0 & 0 & -E & 0 \\
                     0 & -F & 0 & 0 \\
                   \end{array}
                 \right).
   \end{equation}
   The Gaussian detect operator can be chosen with a CM of the form (\ref{sp1}), with
  \begin{eqnarray}
     \gamma_{1}=diag(M_{1},M_{2},M_{1},M_{2}),\label{we2a3}\\
     \gamma_{2}=diag(M_{3},M_{4},M_{3},M_{4}),\label{we2a4}\\
     \gamma_{3}=\left(
                   \begin{array}{cccc}
                     M_5 & 0 & 0 & 0 \\
                     0 & 0 & 0 & -M_6 \\
                     0 & 0 & -M_5 & 0 \\
                     0 & -M_6 & 0 & 0 \\
                   \end{array}
                 \right)\label{we2a5}.
   \end{eqnarray}
  We have $\omega=-\frac{1}{2}diag(\frac{M_{5}^2}{M_{3}},\frac{M_{6}^2}{M_{4}},\frac{M_{5}^2}{M_{3}},\frac{M_{6}^2}{M_{4}})$. Now $\gamma_{1}$ and $\omega$ are diagonalized already. The conditions on $M_{i}$ are (\ref{we2a}) and (\ref{we2b}). Proposition \ref{proposition1} leads to
  \begin{equation}\label{we2a6}
    (A-E)(A-F)\geq 1
  \end{equation}
  for the necessary condition of the separability of Werner-Wolf symmetric $2\times2$ state. In fact the separable condition (\ref{we2a6}) is also a sufficient condition as we will show in section \ref{sec6}. The condition (\ref{we2a6}) is a new result. Numerical calculation shows that the method in \cite{Zhang} is consistent  with (\ref{we2a6}). But analytical derivation of (\ref{we2a6}) from \cite{Zhang} is not clear.

The witness obtained from integral equation of characteristic function is severely restricted by the condition that $\gamma_{M}\in \Omega$. The attempt to lift this condition leads to the following sections of numeric calculation and the connection of witness based criterion with Werner-Wolf criterion. Instead of obtaining the maximal mean of Gaussian operator over product states, Cauchy-Schwartz inequality provides an upper bound of the mean (see Appendix B), which had been used in \cite{Zhang}.

\section{Gaussian witness based on numeric calculation}\label{sec4}

 Numerical calculation in Fock space shows (see subsection \ref{Subsection 4.1}) that for a two-mode Gaussian detect operator $\hat{M}$ with CM, $\gamma_{M}$, has the form of (\ref{we2a0}), the maximal mean of $\hat{M}$ over product states is achieved by the product of single-mode pure Gaussian states. More concretely, the single-mode pure Gaussian state is either vacuum or a squeezed state. Let the product of squeezed state be $|\psi\rangle=\hat{S}|00\rangle$, where $\hat{S}$ is the product of local squeezing operators. Let $\hat{M'}=\hat{S}^{\dagger}\hat{M}\hat{S}$, $|a'\rangle|b'\rangle=\hat{S}^{\dagger}|a\rangle|b\rangle$, then $\langle\hat{M}\rangle_{|ab\rangle}=\langle\hat{M'}\rangle_{|a'b'\rangle}$. The maximal mean of the Gaussian detect operator over product states is invariant up to local squeezing acting on the detect operator. If for some local squeezing operator $\hat{S}$, the maximal mean of $\hat{M'}$ is achieved by $|00\rangle$, then we can judge that the maximal mean of $\hat{M}$ is achieved by the product of local squeezed states. It is more evident to find $|00\rangle$ as the optimal state than a product squeezed state in numerical calculation. What left is how to find $\hat{S}$. Let the associate symplectic transformation of  the unitary transformation $\hat{S}$ be $S=diag(\sqrt{x},\frac{1}{\sqrt{x}},\sqrt{y},\frac{1}{\sqrt{y}})$, then the CM of $\hat{M'}$ is $\gamma_{M}'=S^{-1}\gamma_{M}S^{-T}$. We have $\langle00|\hat{M'}|00\rangle=\frac{4}{\sqrt{det(\gamma_{M}'+I_{4})}}=\frac{4}{\sqrt{\det[\gamma_{M}+diag(x,\frac{1}{x},y,\frac{1}{y})]}}$. It is shown (in subsection \ref{Subsection 4.1}) that $S$ minimizes $\det(\gamma_{M}'+I_{4})$, so $S$ can be determined. Hence we have the desired result $\Lambda=\max_{\hat{\rho}_{s}}Tr(\hat{\rho}_{s}\hat{M})=\frac{4}{\min_{x,y}\sqrt{\det[\gamma_{M}+diag(x,\frac{1}{x},y,\frac{1}{y})]}}$. Then it follows
\begin{proposition}\label{proposition2}
For a two-mode Gaussian state with standard\cite{Duan}\cite{Simon} CM, $\gamma$, the necessary criterion of separability is
\begin{equation}\label{we1q}
   \mathcal{L}=\min_{\gamma_{M}}\frac{\det(\gamma+\gamma_{M})}{\min_{x,y}\det(diag(x,\frac{1}{x},y,\frac{1}{y})+\gamma_{M})}\geq 1,
\end{equation}
for $\gamma_{M}$ given in (\ref{we2a0}) and $\gamma_{M}\geq i\sigma $.
\end{proposition}
The condition $\gamma_{M}\geq i\sigma$ guarantees the positivity of $\hat{M}$.

\subsection{Method and result of numeric calculation}\label{Subsection 4.1}

\begin{figure}[tpp]
\includegraphics[ trim=0.000000in 0.000000in 0.000000in 0.500000in, width=3.5in]{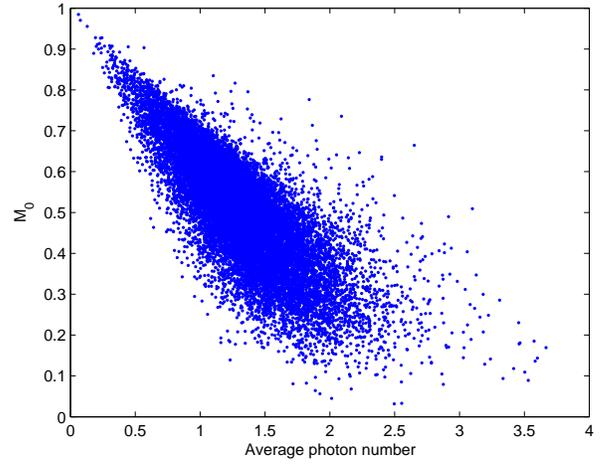}
\caption{The calculation of (\ref{sp9}) with random generated Gaussian detect operator $\hat{M}$ and random state vector $|\psi_{1}\rangle$. The average photon number is the average of mean photon numbers in $|\psi_{1}\rangle$ and $|\psi_{2}\rangle$. The lengths of vectors $|\psi_{i}\rangle$ are set to be 6 as the cutoff.}
\label{Fig1}
\end{figure}

We will illustrate the method of numeric calculation of the maximal mean of a Gaussian detect operator over product state set.
We consider the CM of a Gaussian detect operator $\hat{M}$ with the form of (\ref{we2a0}). The matrix elements of $\hat{M}$ in Fock space is \cite{Chen07}
\begin{eqnarray}\label{sp3}
&&\hat{M} _{k_1,k_2;m_1,m_2}=\tilde{c}(\partial
t_1)^{k_1}(\partial t_1^{\prime })^{m_1}(\partial t_2)^{k_2}(\partial
t_2^{\prime })^{m_2}  \nonumber \\
&&\times\exp \{\frac 12(t,t^{\prime })[\sigma _1\otimes I_2+\beta ](t,t^{\prime
})^T\}\left| _{t_{_1}, t_2,t_1^{\prime }, t_2^{\prime
}=0}\right.
\end{eqnarray}
where $t=(t_1,t_2),$ $t^{\prime }=(t_1^{\prime },t_2^{\prime
});$ $(\partial t)^k$ stands for $\partial ^k/\partial
t^k;\sigma _i$ are the Pauli matrices. The normalization factor $\tilde{c}=\sqrt{\frac{\det(\beta)}{k_{1}!k_{2}!m_{1}!m_{2}!}}$. The matrix $\beta $ is
\begin{equation}\label{sp4}
\beta =(\sigma _3\otimes I_2)(\tilde{\gamma}_M/2 +\sigma _1\otimes I_2/2)^{-1}(\sigma
_3\otimes I_2),
\end{equation}
where $\tilde{\gamma}_M$ is the complex covariance matrix of $\hat{M}$, it is a transform of CM and the definition can be found in next section. With matrix $\sigma _1\otimes I_2+\beta$, we can directly evaluate the maximal mean of $\hat{M}$ over product pure states. The numerical calculation shows that the product state $|\psi\rangle=|\psi_{1}\rangle|\psi_{2}\rangle$ maximize the quantity $\langle\psi|\hat{M}|\psi\rangle$ is with squeezed states $|\psi_{1}\rangle$ and $|\psi_{2}\rangle$. Due to the cutoff of the length of vectors $|\psi_{i}\rangle$ $(i=1,2)$ in numeric calculation, the numeric extremal states $|\psi_{i}\rangle$ are approximate squeezed states. A more strict and apparent way is to use the pre-squeezed Gaussian detect operator $\hat{M}'=\hat{S}^{\dagger}\hat{M}\hat{S}$. The symplectic transformation $S$ can be found by minimizing $\det(\gamma_M+diag(x,\frac{1}{x},y,\frac{1}{y}))$ with respect to $x,y$. It leads to equations for $x,y$:

\begin{eqnarray}
  &&(M_{3}+y)[M_{2}(M_{4}+\frac{1}{y})-M_{6}^2]\nonumber\\
  &&-\frac{1}{x^2}(M_{4}+\frac{1}{y})[M_{1}(M_{3}+y)-M_{5}^2]=0 \label{sp5}, \\
  &&(M_{1}+x)[M_{4}(M_{2}+\frac{1}{x})-M_{6}^2]\nonumber\\
  &&-\frac{1}{y^2}(M_{2}+\frac{1}{x})[M_{3}(M_{1}+x)-M_{5}^2]=0 \label{sp6}.
\end{eqnarray}
On the other hand, we denote $\tilde{\gamma}_M+\sigma_{1}\otimes I_{2}=\left( \begin{array}{cc} U& V \\ V & U \\ \end{array}\right)$,
where $U=\left( \begin{array}{cc} K_{1}& K_{3} \\ K_{3} & K_{5}\\ \end{array}\right)$, $V=\left( \begin{array}{cc}K_{2}& K_{4}\\K_{4} & K_{6} \\ \end{array}\right)$, with $ K_{1}=\frac{1}{2}(M_{2}x-M_{1}/x)$, $ K_{2}=\frac{1}{2}(M_{2}x+M_{1}/x)+1$,  $K_{3}=\frac{1}{2}(M_{4}y-M_{3}/y)$,
$K_{4}=\frac{1}{2}(M_{4}y+M_{3}/y)+1$, $K_{5}=-\frac{1}{2}(\sqrt{xy}M_{6}+\frac{1}{\sqrt{xy}}M_{5})$, $K_{6}=\frac{1}{2}(-\sqrt{xy}M_{6}+\frac{1}{\sqrt{xy}}M_{5})$. Denoting $T=(U-VU^{-1}V)^{-1}$,
the matrix $\sigma _1\otimes I_2+\beta $ in (\ref{sp3}) then is
\begin{equation}
  \left(
    \begin{array}{cc}
      2T& 1+2U^{-1}VT\\
      1+2U^{-1}VT& 2T\\
    \end{array}
  \right).
\end{equation}


The diagonal elements of $T$ are $K_{1}-\frac{K_{3}K_{2}^2+K_{1}K_{6}^2-2K_{2}K_{5}K_{6}}{K_{1}K_{3}-K_{5}^2}$, $ K_{3}-\frac{K_{3}K_{6}^2+K_{1}K_{4}^2-2K_{4}K_{5}K_{6}}{K_{1}K_{3}-K_{5}^2}$, respectively. They are equal to zeros due to Eqs. (\ref{sp5}) and (\ref{sp6}). The matrix $\sigma _1\otimes I_2+\beta $ then can be expressed as
\begin{equation}\label{sp7}
 \left(
   \begin{array}{cccc}
     0 & N_{1} & N_{2} & N_{3} \\
     N_{1} & 0& N_{3} & N_{4} \\
     N_{2} & N_{3} & 0 & N_{1} \\
     N_{3} & N_{4} & N_{1} & 0 \\
   \end{array}
 \right),
\end{equation}
where $N_{i}$ $(i=1,...,4)$ can be worked out explicitly as the functions of $K_{j}$ $(j=1,...,6)$ and at last they are functions of $M_{i}$ $(i=1,...,6)$, Notice that the optimal $x,y$ appeared in $K_{j}$ are also determined by $M_{i}$ . Hence
\begin{eqnarray}\label{sp8}
 \exp \{\frac 12(t,t^{\prime })[\sigma _1\otimes I_2+\beta ](t,t^{\prime})^T\}\nonumber\\
=\sum_{k,l,m,n,i,j=0}^{\infty}\frac{N_{1}^{i+j}N_{2}^{k}N_{3}^{m+n}N_{4}^{l}}{k!l!m!n!i!j!}\nonumber\\
\times t_{1}^{k+m+i}t_{2}^{l+n+i}t_{1}^{\prime k+n+j}t_{2}^{\prime l+m+j}.
\end{eqnarray}

Let the product state be $|\psi\rangle=|\psi_{1}\rangle|\psi_{2}\rangle$, with the normalized $|\psi_{1}\rangle=\sum_{k=0}^{\infty}a_{k}|k\rangle, |\psi_{2}\rangle=\sum_{k=0}^{\infty}b_{k}|k\rangle$. Then $\langle\hat{M}\rangle_{|\psi\rangle}=\sum_{k_1,k_2,m_1,m_2}a_{k_1}^{*}b_{k_2}^{*}\hat{M}_{k_1,k_2;m_1,m_2}a_{m_1}b_{m_2}$. We denote $\langle\hat{M}\rangle_{|\psi\rangle}=\sqrt{\det(\beta)}M_{0}$, with $M_{0}$ defined as
\begin{equation}\label{sp9}
\sum_{k,l,m,n,i,j}a_{k_1}^{*}b_{k_2}^{*}a_{m_1}b_{m_2}\frac{N_{1}^{i+j}N_{2}^{k}N_{3}^{m+n}N_{4}^{l}\sqrt{k_1!k_2!m_1!m_2!}}{k!l!m!n!i!j!},
\end{equation}
where $k_1=k+m+i,k_2=l+n+i,m_1=k+n+j,m_2=l+m+j$. Numerical calculation shows that the maximal $M_{0}$ is achieved when $a_{0}=b_{0}=1$ as far as $\gamma_{M}\geq0$, so that
\begin{eqnarray}\label{sp9b}
&  \max_{|\psi\rangle}\langle\hat{M}\rangle=\sqrt{\det\beta}\nonumber\\
&=\frac{4}{\min_{x,y}\sqrt{\det[\gamma_{M}+diag(x,1/x,y,1/y)]}}.
\end{eqnarray}

For given $\hat{M}$ and $|\psi_{2}\rangle$, the mean of $\hat{M}$ over $|\psi_{2}\rangle$ is a matrix, namely, $\langle \psi_{2}|\hat{M}|\psi_{2}\rangle$. This matrix has a largest eigenvalue with some eigenvector. We choose $|\psi_{1}\rangle$ to be equal to the eigenvector corresponding to the largest eigenvalue. The problem of the maximal mean of $\hat{M}$ over product states is reduced to the problem of the largest eigenvalue of matrix $\langle \psi_{2}|\hat{M}|\psi_{2}\rangle$. We have calculated the largest eigenvalue and the corresponding eigenvector for the  matrix. The results of numerical calculation of $M_{0}$ are shown in Fig.\ref{Fig1}. There is a statistical tendency that $M_{0}$ decreases when the average photon number increases. When the average photon number tends to zero (the product vacuum state), $M_{0}$ tends to one and we may anticipate that (\ref{sp9b}) is true.

For equation (\ref{sp9b}), further numeric evidence comes from the iterative calculation. In the calculation, if $M_{1}M_{2}\leq M_{3}M_{4}$ (otherwise we interchange $M_{1}$ with $M_{3}$ and $M_{2}$ with $M_{4}$ or interchange the roles of $|\psi_{1}\rangle$ and $|\psi_{2}\rangle$), we input a random vector $|\psi_{2}\rangle$, calculate the largest eigenvalue of $\langle \psi_{2}|\hat{M}|\psi_{2}\rangle$ and the corresponding eigenvector $|\psi_{1}\rangle$. In the next round of iteration, we replace $|\psi_{2}\rangle$ with $|\psi_{1}\rangle$.  The iterative ends up with $|\psi_{1}\rangle$ and $|\psi_{2}\rangle$ tending to vacuum and $M_{0}\rightarrow 1$. It takes several rounds of iterations to reach $M_{0}>0.99999$ (The statistic shows the number of rounds has a peak at 4 and an average at about 7).  For each round of iteration, we map the input state $|\psi_{2}\rangle$ to output state $|\psi_{1}\rangle$. We have observed that the mean photon number of output is less than that of input. Notice that every iteration successes in the sense that it decrease the mean photon number while increase the quantity $M_{0}$. A possible explanation of the success is that in the map, the input state $|\psi_{2}\rangle$ annihilates the mode in $\hat{M}$ with more photon (specified with larger $M_{3}M_{4}$) and leave the mode with less photon (specified with smaller $M_{1}M_{2}$). One the other hand, if $M_{1}M_{2}\leq M_{3}M_{4}$, we input a random vector $|\psi_{1}\rangle$ to obtain the matrix $\langle \psi_{1}|\hat{M}|\psi_{1}\rangle$. Then we choose $|\psi_{2}\rangle$ to be the eigenvector of the matrix $\langle \psi_{1}|\hat{M}|\psi_{1}\rangle$ corresponding to the largest eigenvalue. Although numeric calculation shows that the iteration converges to $M_{0}=1$ and $|\psi\rangle=|00\rangle$ in many cases, there are also cases that the iteration does not converge.

\subsection{Two mode squeezed thermal state}

The two mode squeezed thermal state is characterized by its CM in (\ref{we2}) with $\mathcal{A}=aI_{2} , \mathcal{B}=bI_{2}, \mathcal{C}=c\sigma_{3} $. Let $\gamma_{M}$ be the form of (\ref{we2a0}), with $M_{2}=M_{1}, M_{4}=M_{3}, M_{6}=M_{5}$. Then $x=1,y=1$ is the solution in minimizing $\det(\gamma_{M}+diag(x,\frac{1}{x},y,\frac{1}{y}))$. Assuming $a\geq b$ without loss of generality, we further let $M_{5}^2=(M_{1}+1)(M_{3}-1)$, then
\begin{eqnarray}
  &\mathcal{L}=\frac{(M_{1}+a)(M_{3}+b)-(M_{5}+c)^2}{(M_{1}+1)(M_{3}+1)-M_{5}^2}=\frac{(a-1)(b+1)-c^2}{2(M_{1}+1)}\nonumber \\
  &+\frac{(M_{3}-1)(a-1)+(M_{1}+1)(b+1)-2M_{5}c}{2(M_{1}+1)}.
\end{eqnarray}
The first term can be omitted when $M_{i}\rightarrow\infty$. The second term can be minimized with respect to the $\sqrt{\frac{M_{3}-1}{M_{1}-1}}$. When $M_{i}\rightarrow\infty$, we have $\mathcal{L}\rightarrow \frac{1}{2}(b+1)-\frac{c^2}{2(a-1)}\geq 1$, which is the necessary condition of separability
\begin{equation}\label{sp9c}
  (a-1)(b-1)\geq c^2.
\end{equation}
The condition is also sufficient\cite{Duan}\cite{Simon}.
\section{Non-Gaussian entanglement}\label{sec5}
The entanglement detecting power of the witness constructed from a Gaussian detect operator $\hat{M}$ discussed so far is limited to Gaussian states. The entanglement of a non-Gaussian state prepared by adding photon to or/and subtracting photon from a Gaussian state (NGPASG) can also be detected with witness based on a Gaussian detect operator $\hat{M}$. We consider an $\hat{M}$ with zero mean and CM $\gamma_{M}$. Then, its characteristic function possesses the property of  $\chi_{M}(\mu)=\chi_{M}(-\mu)$.  It is more convenient to deal with the annihilation operator and creation operator directly for a photon added or subtracted state. Therefore, we need to transform a CM $\gamma$ to a complex covariance matrix (CCM) $\tilde{\gamma}$, which is given by
  $\tilde{\gamma}=\frac{1}{2}\left(
  \begin{array}{cc}
  \gamma^p-\gamma^x+i(\gamma^{xp}+\gamma^{px}) & \gamma^p+\gamma^x+i(\gamma^{xp}-\gamma^{px}) \\
  \gamma^p+\gamma^x-i(\gamma^{xp}-\gamma^{px})& \gamma^p-\gamma^x-i(\gamma^{xp}+\gamma^{px}) \\
   \end{array}
   \right)
  $. When the order of quadrature operators in $\hat{R}$ is properly arranged, the CM $\gamma$ of a state can be rewritten as $\left(
                                                                                                   \begin{array}{cc}
                                                                                                     \gamma^x & \gamma^{xp} \\
                                                                                                     \gamma^{px} & \gamma^p \\
                                                                                                   \end{array}
                                                                                                 \right)
$, with $\gamma^x_{i,j}=\gamma_{2i-1,2j-1}, \gamma^p_{i,j}=\gamma_{2i,2j}$ and $\gamma^{xp}_{i,j}=\gamma^{px}_{j,i}=\gamma_{2i-1,2j}=\gamma_{2j,2i-1}$.
The characteristic function of a zero-mean Gaussian state then is $\chi(\mu)=Tr[\hat{\rho}\mathcal{D}(\mu)]=\exp[-\frac{1}{4}(\mu,\mu^*)\tilde{\gamma}(\mu,\mu^*)^T]$, where $\mathcal{D}(\mu)=\exp[\Sigma_{j=1}^{n}(\mu_j\hat{a}_{j}^{\dagger}-\mu_{j}^*\hat{a}_{j})]$ is the displacement operator, and $\hat{a}_{j},\hat{a}_{j}^{\dagger}$ are the annihilation and creation operators of the $j$-th mode, respectively. We consider a non-Gaussian state prepared with photon addition or/and subtraction, and the state can be written as $\hat{\rho}=\mathcal{N}\hat{a}^{k\dagger}\hat{a}^{m}\hat{\rho}_{G}\hat{a}^{m\dagger}\hat{a}^{k}$, where $\hat{\rho}_{G}$ is a zero-mean Gaussian state (we call it Gaussian kernel of state $\hat{\rho}$ ) with CM, $\gamma_{G}$ (CCM $\tilde{\gamma}_{G}$), $\mathcal{N}$ is the normalization. The characteristic function of the kernel is denoted as $\chi_{G}$. Herein $\hat{a}^k$ is a short notation of $\hat{a}_{1}^{k_{1}}\hat{a}_{2}^{k_{2}}...\hat{a}_{n}^{k_{n}}$ and similar notation for the creation operator. The non-Gaussian state can be generated from Gaussian kernel operator function\cite{Jiang14},
  \begin{equation}\label{we8}
    \hat{Q}(\varepsilon,\xi,\eta,\zeta)=e^{\varepsilon\hat{a}^{\dagger}}e^{\xi \hat{a}}\hat{\rho}_{G}e^{\eta\hat{a}^{\dagger}}e^{\zeta\hat{a}},
  \end{equation}
   by derivatives. Namely, $\hat{\rho}=\mathcal{N}\mathcal{O}\hat{Q}(\varepsilon,\xi,\eta,\zeta)$, where $\varepsilon \hat{a}^{\dagger}=\sum_{j=1}^{n}\varepsilon_{j} \hat{a}_{j}^{\dagger}$ and $\mathcal{O}=\frac{\partial^{2|k|+2|m|}}{\partial\varepsilon^{k}\partial\xi^{m}\partial\eta^{m}\partial\zeta^{k}}|_{\varepsilon=\xi=\eta=\zeta=0}$, with $|k|=\sum_{j=1}^{n}k_{j}$ and $\partial\varepsilon^{k}=\partial\varepsilon_{1}^{k_{1}}\partial\varepsilon_{2}^{k_{2}}...\partial\varepsilon_{n}^{k_{n}}$. Here $\varepsilon,\xi,\eta,\zeta$ are real vectors.  The characteristic function of operator $\hat{Q}(\varepsilon,\xi,\eta,\zeta)$ is
  \begin{eqnarray}\label{we9}
   &\chi_{Q}(\mu,\varepsilon,\xi,\eta,\zeta)=\chi_{Q}(0,\varepsilon,\xi,\eta,\zeta)\chi_{G}(\mu) \nonumber\\
   &\times\exp[\frac{1}{2}(-\varepsilon,\zeta)\tilde{\gamma}_{G_{+}}(\mu,\mu^*)^T+\frac{1}{2}(-\eta,\xi)\tilde{\gamma}_{G_{-}}(\mu,\mu^*)^T],
  \end{eqnarray}
  where $\tilde{\gamma}_{G_{\pm}}=\tilde{\gamma}_{G}\pm \sigma_{1}\otimes I_{n}$ and
  \begin{eqnarray}
    &&\chi_{Q}(0,\varepsilon,\xi,\eta,\zeta)=\exp[-\frac{1}{4}(\varepsilon,-\zeta)\tilde{\gamma}_{G_{+}}(\varepsilon,-\zeta)^T \nonumber\\
    &&-\frac{1}{2}(\varepsilon,-\zeta)\tilde{\gamma}_{G_{-}}(\eta,-\xi)^T-\frac{1}{4}(\eta,-\xi)\tilde{\gamma}_{G_{-}}(\eta,-\xi)^T]. \nonumber
  \end{eqnarray}
  In deriving of (\ref{we9}), we have used coherent state representation by inserting several identities of $\int|\alpha\rangle\langle\alpha|[\frac{d^{2}\alpha}{\pi}]\equiv1$ (where $\alpha=(\alpha_{1},...,\alpha_{n})$). Properly arranging the order of integrals  leads to (\ref{we9}).

  We thus have
  \begin{eqnarray}\label{we10}
     && Tr(\hat{\rho} \hat{M})=\mathcal{N}\mathcal{O}Tr[\hat{Q}(\varepsilon,\xi,\eta,\zeta)\hat{M}]\nonumber\\
     &&=\mathcal{N}\mathcal{O}\int[\frac{d^2\mu}{\pi}]\chi_{Q}(\mu,\varepsilon,\xi,\eta,\zeta)\chi_{M}(\mu)\nonumber\\
     && =\frac{2^n\mathcal{N}\mathcal{O}\chi_{Q}(0,\varepsilon,\xi,\eta,\zeta)\exp[f(\varepsilon,\xi,\eta,\zeta)]}{\sqrt{|\det(\tilde{\gamma}_{G}+\tilde{\gamma}_{M})|}}
  \end{eqnarray}
  with $f(\varepsilon,\xi,\eta,\eta)=\frac{1}{4}[(\varepsilon,-\zeta)\tilde{\gamma}_{G_{+}}+(\eta,-\xi)\tilde{\gamma}_{G_{-}}](\tilde{\gamma}_{G}+\tilde{\gamma}_{M})^{-1}[\tilde{\gamma}_{G_{+}}(\varepsilon,-\zeta)^T+\tilde{\gamma}_{G_{-}}(\eta,-\xi)^T]$. In the limit of $\tilde{\gamma}_{M}\rightarrow\infty$ (also denoted as $\gamma_{M}\rightarrow\infty$), that is, all the parameters, $M_{i}$, in the CM $\gamma_{M}$ tend to infinite or negative infinite, we have $f(\xi,\eta)\rightarrow 0$. Further notice that $Tr(\hat{\rho})=\chi_{\hat{\rho}}(0)=\mathcal{N}\mathcal{O}\chi_{Q}(0,\varepsilon,\xi,\eta,\zeta)=1$. Then
  \begin{equation}\label{we11}
    Tr(\hat{\rho} \hat{M})_{\gamma_{M}\rightarrow\infty}\rightarrow \frac{2^n}{\sqrt{|\det(\tilde{\gamma}_{G}+\tilde{\gamma}_{M})|}}=\frac{2^n}{\sqrt{|\det(\gamma_{G}+\gamma_{M})|}}.
  \end{equation}
  Then we have the NGPASG version of Proposition \ref{proposition1}:
  \begin{proposition}\label{proposition3}
  The necessary criterion of separability for a photon added (subtracted) Gaussian state is
  \begin{equation}\label{we12}
   \det(\gamma_{G}+\gamma_{M})\geq \det(\gamma_{A}\oplus\gamma_{B}+\gamma_{M}),
\end{equation}
for $\gamma_{M}\in\Omega$ and $\gamma_{M}+i\sigma\geq 0$ with infinite parameters $M_{i}$. Herein $\gamma_{G}$ is the CM of Gaussian kernel of photon added (subtracted) state $\hat{\rho}$, $\gamma_{A}$ and $\gamma_{B}$ are the CMs of the subsystem pure Gaussian states.
\end{proposition}

The criterion (\ref{we12}) comes from (\ref{we11}) and $Tr(\hat{\rho} \hat{M})\leq \Lambda$ if $\hat{\rho}$ is separable, with $\Lambda=2^n[\det{(\gamma_{A}\oplus\gamma_{B}+\gamma_{M})}]^{-\frac{1}{2}}$ has already been determined for a given $\gamma_{M}\in\Omega$ in the derivation of criterion (\ref{we1}).

The criterion (\ref{we12}) is almost the same as criterion (\ref{we1}) except for the condition that $\gamma_{M}\rightarrow\infty$. It means that the necessary criterion of separability for a photon added (subtracted) state can be reduced to the necessary criterion of separability of its Gaussian kernel under the condition of infinite $\gamma_{M}$.

If the kernel state is a two mode symmetric Gaussian state, the necessary condition (\ref{we2a1}) for its separability is sufficient. Then (\ref{we2a1}) is also the necessary and sufficient condition for the photon added or subtracted states prepared from the two mode symmetric Gaussian state. The sufficiency of NGPASG comes from the fact that addition and subtraction of photon(s) are local operations. The special case of necessary and sufficient condition of separability for a two-mode NGPASG prepared by a single photon adding to (subtracting from) a symmetric Gaussian state at each mode had been shown in \cite{Jiang14}.

The NGPASG version of Proposition \ref{proposition2} is as follows due to (\ref{we11}):
\begin{proposition}\label{proposition4}
For a $1\times1$ NGPASG state $\hat{\rho}=\mathcal{N}\hat{a}_{1}^{\dagger k_{1}}\hat{a}_{2}^{\dagger k_{2}}\hat{a}_{1}^{m_{1}}\hat{a}_{2}^{m_{2}}\hat{\rho}_{G}\hat{a}_{1}^{\dagger m_{1}}\hat{a}_{2}^{\dagger m_{2}}\hat{a}_{1}^{k_{1}}\hat{a}_{2}^{k_{2}}$ with Gaussian kernel $\hat{\rho}_{G}$ characterized by CM, $\gamma_{G}$, the necessary criterion of separability is
\begin{equation}\label{we13}
  \mathcal{L}=\min_{\gamma_{M}}\frac{\det(\gamma_{G}+\gamma_{M})}{\min_{x,y}\det(diag(x,1/x,y,1/y)+\gamma_{M})}\geq 1,
\end{equation}
regardless the number of photon added to or subtracted from the Gaussian kernel. Where $M_{i}\rightarrow\infty$ and $\gamma_{M}\geq i\sigma$.
\end{proposition}

The necessary condition (\ref{sp9c}) for a two mode squeezed thermal state is also sufficient, so we have the necessary and sufficient condition (\ref{sp9c}) for the separability of a state prepared by adding or/and subtracting any number of photons to (from) a two mode squeezed thermal state. The sufficiency comes from the fact that addition and subtraction of photon(s) are local operations.

\begin{figure}[btp]
\centering
\subfigure[\label{Fig.2a}]{
\includegraphics[width=1.65in]{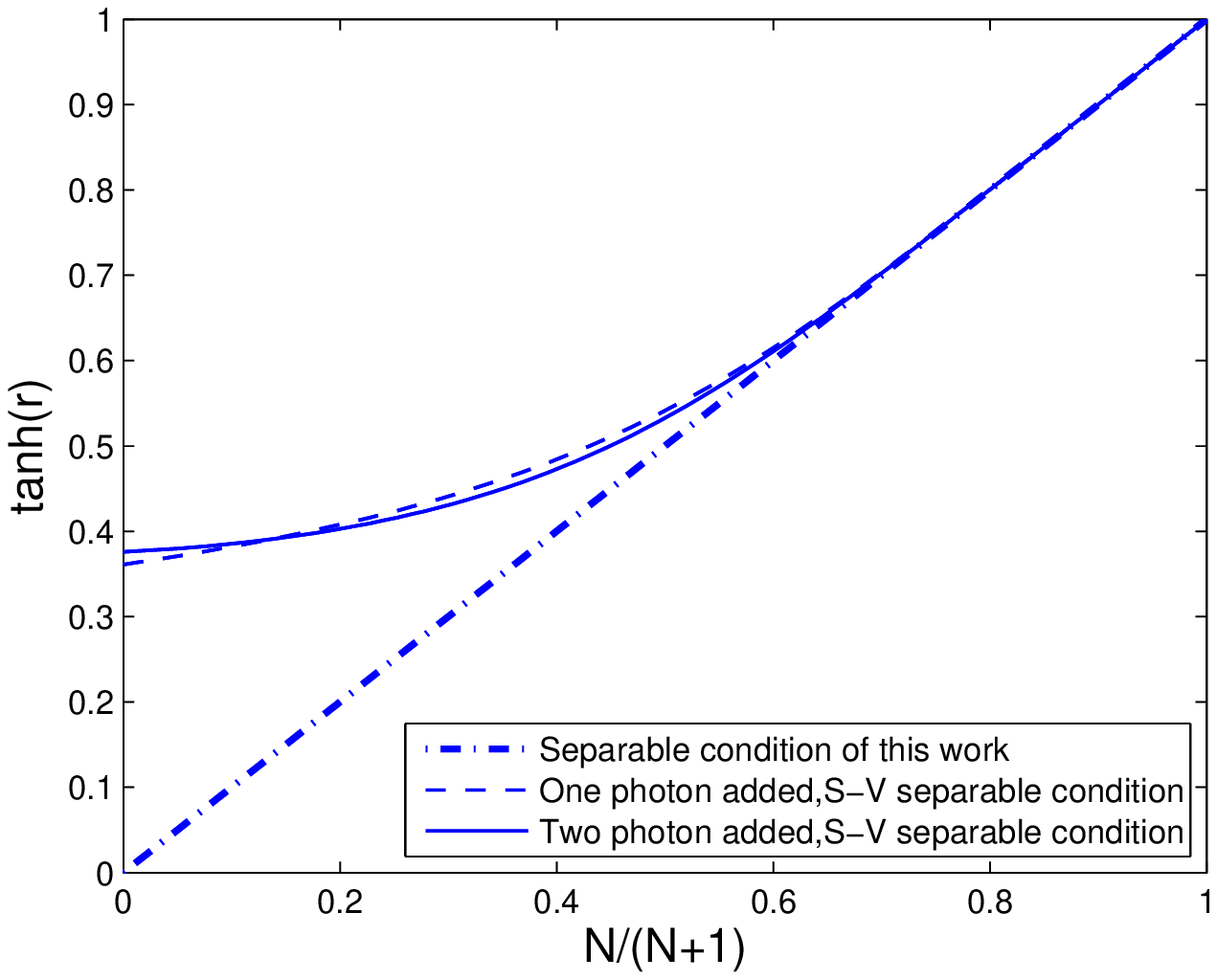}}
\subfigure[\label{Fig.2b}]{
\includegraphics[width=1.65in]{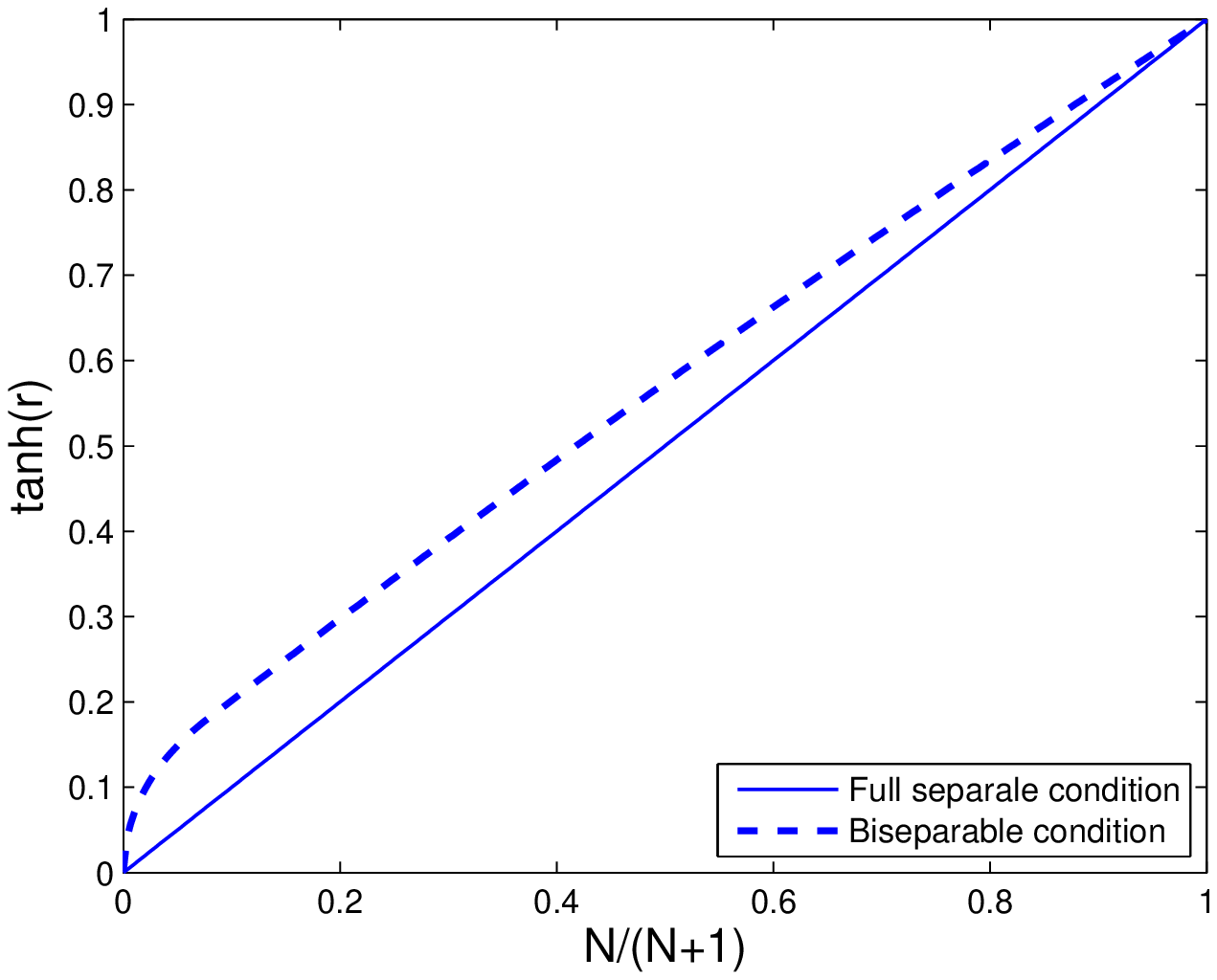}}
\caption{(a)Photon added two mode symmetric squeezed thermal state. Necessary and sufficient separable criterion of this work and S-V necessary criterion \cite{Shchukin}. The one photon added and two photon added states of this work share the same condition of separability, $\tanh(r)\leq \frac{N}{N+1}$.
(b)Three mode symmetric squeezed thermal state. $a=(2N+1)\frac{1}{3}(e^{2r}+2e^{-2r}), c=(2N+1)\frac{1}{3}(e^{2r}-e^{-2r})$. From top left to bottom right: genuine entangled, biseparable, full separable.}
\label{Fig2}
\end{figure}

When the kernel Gaussian state is the two mode symmetric squeezed thermal state described by CM in (\ref{we2}), with $\mathcal{A}=\mathcal{B}=(2N+1)\cosh(2r)I_{2}$ and $\mathcal{C}=(2N+1)\sinh(2r)\sigma_{3}$. The conditions (\ref{we2a1}) and (\ref{sp9c}) are equivalent.
A comparison of our separability condition (\ref{we2a1}) and criterion in \cite{Shchukin} for NGPASG is shown in Fig.\ref{Fig2}. It is evident that our criterion is better than the criterion in \cite{Shchukin} is in detecting entanglement of photon added NGPASG.

\section{Refined Werner-Wolf criterion}\label{sec6}
The applications of necessary criterion of separability in (\ref{we1}) are limited by the additional condition of $\gamma_{M}\in \Omega$. In fact, we can get rid of the condition $\gamma_{M}\in \Omega$ to obtain the following result.

\begin{proposition}\label{proposition5}
The necessary criterion of separability for a Gaussian state with CM $\gamma$ is
\begin{equation}\label{we2e}
   \det(\gamma+\gamma_{M})\geq \det(\gamma_{A}\oplus\gamma_{B}+\gamma_{M}),
\end{equation}
for any given positive $\gamma_{M}$, with $\det(\gamma_{A})=\det(\gamma_{B})=1$.
\end{proposition}

The proof comes from the Werner-Wolf necessary and sufficient criterion of separability for a Gaussian state, $\gamma\geq\gamma_{A}'\oplus\gamma_{B}'$ \cite{Werner}, where $\gamma_{A}', \gamma_{B}'$ are CMs of the subsystem states that may not be pure. Any CM $\gamma_{j}'$ $(j=A$ or $B)$ can be transformed by some sympletic transform $S$ into diagonal form $V =S\gamma_{j}'S^{T}$ with symplectic eigenvalues $v_{i}\geq 1$ of multiplicity 2 as diagonal elements. Clearly, we have $V\geq I$, where identity matrix $I$ characterizes vacuum state of $n_{j}$ modes. Then, we have $\gamma_{j}'\geq S^{-1} (S^{T})^{-1}\equiv \gamma_{j}$, since for any $2n_{j}$ dimensional vector $u$, we have $uS^{-1}(V-I)(S^{-1})^Tu^T\geq0$ due to the nonnegativity of $V-I$. Therefore, we have $S^{-1}(V-I)(S^{-1})^T\geq 0$, and $\gamma_{j}'\geq \gamma_{j}$ follows. The symplectic transform $S$ should keep $\sigma^{-1}$ invariant, $S\sigma^{-1}S^T=\sigma^{-1}$, so that $\det(SS^T)=1$. Then, $\gamma_{j}$ is the CM of a pure Gaussian state. We have the refined Werner-Wolf necessary and sufficient criterion of separability, which is
\begin{equation}\label{we2f}
  \gamma\geq\gamma_{A}\oplus\gamma_{B},
\end{equation}
with pure states in the subsystems. We may add a positive $\gamma_{M}$ to both sides of the refined Werner-Wolf criterion, take determinants and minimizing the right hand side of the inequality with respect to $\gamma_A$ and $\gamma_B$ to complete the proof of the proposition. Notice that $\gamma_A$ and $\gamma_B$ in Proposition {\ref{proposition5}} do not rely on $\gamma_M$.

\subsection{Werner-Wolf $2\times2$ state}
The CM of Werner-Wolf bounded entangled $2\times2$ state \cite{Werner} (with two modes at both parties) can be extended to a more general form of (The state then is called generalized Werner-Wolf state) (\ref{we2}) with $\mathcal{A}=diag(A,B,A,B)$,$\mathcal{B}=diag(C,D,C,D)$ and $\mathcal{C}$ takes the form of (\ref{we2a2}).

We then choose $\gamma_{M}$ to be the same structure of $\gamma$ with $A$,...,$F$ substituted by $M_{1}$,...,$M_{6}$, respectively. Namely, $\gamma_{M}$ is determined by (\ref{sp1}) with (\ref{we2a3})(\ref{we2a4})(\ref{we2a5}). Assuming $\gamma_{A}=diag(1/x,x,1/x,x),\gamma_{B}=diag(y,1/y,y,1/y)$, then the separable criterion of (\ref{we2f}) reduces to
\begin{eqnarray}
  (A-\frac{1}{x})(C-\frac{1}{y})\geq E^2  \label{we2f1} \\
  (B-x)(D-y)\geq F^2 \label{we2f2}
\end{eqnarray}
The condition for the existence of at least a pair $(x,y)$ solution for inequalities (\ref{we2f1}) and (\ref{we2f2}) can be derived as
\begin{equation}\label{we5a}
  (AC-E^2)(BD-F^2)-2|EF|-CD-AB+1\geq 0.
\end{equation}
It is the explicit necessary criterion of separability for the generalized Werner-Wolf state. The criterion in (\ref{we5a}) is necessary and sufficient for the separability of the generalized Werner-Wolf state, due to the fact that Werner-Wolf criterion is necessary and sufficient for separability. The condition (\ref{we2a6}) is a special case of (\ref{we5a}), so that (\ref{we2a6}) is also a sufficient condition for separability.

 With local operations, the CM of a two-mode Gaussian state can always be put into the standard form \cite{Simon}\cite{Duan} of (\ref{we2}),
 with $\mathcal{A}=diag(a,a)$,$\mathcal{B}=diag(b,b)$,$\mathcal{C}=diag(c_{1},-c_{2})$. Following the same procedure of above treatment for generalized Werner-Wolf state, we have the necessary criterion of separability for the two mode Gaussian state
\begin{equation}\label{we7}
  (ab-c_{1}^2)(ab-c_{2}^2)-2|c_{1}c_{2}|-a^2-b^2+1\geq 0.
\end{equation}
The criterion (\ref{we7}) is just Simon's criterion \cite{Simon}, which is known to be necessary and sufficient for the separability of the two-mode Gaussian state with standard CM.
\subsection{Multi-mode symmetric Gaussian state}
Further examples for the separability of $1\times1\times...\times1$ and $1\times2$ symmetric Gaussian states will be presented  as the applications of criterion (\ref{we2f}).

It is straightforward to extend the refined Werner-Wolf criterion to the necessary criterion of the full separability of multipartite Gaussian states. It reads
\begin{equation}\label{we2c}
  \gamma\geq \gamma_{A_{1}}\oplus\gamma_{A_{2}}\oplus...\oplus\gamma_{A_{n}},
\end{equation}
where the system is split into subsystems $A_{1},...,A_{n}$ for the $n$-partite fully separable problem. The states in the subsystems $A_{j} (j=1,...,n)$ are pure states,namely, $\det(\gamma_{A_{j}})=1, (j=1,...,n) $. The criterion can be applied to the fully separable condition of multipartite continuous variable GHZ state of $n$ modes \cite{Chen05}\cite{vonLoock}.
The CM of the symmetric multi-mode Gaussian state is $\gamma=\gamma^x\oplus\gamma^p$. The diagonal elements of $\gamma^x$ and $\gamma^p$  are $a$ and $b=a+(n-2)c$, respectively. The off-diagonal elements of $\gamma^x$ and $\gamma^p$ are $c$ and $-c$, respectively. Due to the symmetry, the criterion of separability leads to $\gamma^x\geq qI_n$ and $\gamma^p\geq \frac{1}{q}I_n$ for some $q>0$ yet to be determined. We can show that $q=1$, then the necessary criterion of full separability reads: $a-c-1\geq 0$ when $c>0$ and $b+c-1\geq 0$ when $c<0$. For a more general symmetric Gaussian state with CM, $\gamma=\gamma^x\oplus\gamma^p$, the diagonal elements of $\gamma^x$ and $\gamma^p$  are $a$ and $b$, respectively. The off-diagonal elements of $\gamma^x$ and $\gamma^p$ are $c_1$ and $-c_2$, respectively. We assume $c_1>0$ and $c_2>0$, the condition of full separability is
\begin{equation}\label{sp9a}
  (a-c_1)(b-(n-1)c_2)\geq 1.
\end{equation}

For the three mode symmetric Gaussian state with CM specified with above parameters $a,b,c$, where $b=a+c$, and $c>0$ (It can also be called three mode symmetric squeezed thermal state). Let us denote the three modes as $A,B$ and $C$,respectively. To study the genuine entanglement of the state, let us consider a $1\times2$ partition, with $A$ mode in one party and $BC$ modes in the other which is denoted as split $A|BC$. We also have splits $B|AC$ and $C|AB$. The state is biseparable if it can be written as $\rho=p_{1}\rho_{A|BC}+p_{2}\rho_{B|AC}+p_{3}\rho_{C|AB}$, it is genuine entangled otherwise, where $p_{i}$ is the probability and $\rho_{A|BC}$ is a separable state with respect to the partition $A|BC$.  In the CM level, the necessary and sufficient criterion of biseparability for the three mode Gaussian state with CM, $\gamma$, is \cite{Eisert}
\begin{equation}\label{sp9b0}
  \gamma-(p_{1}\gamma_{A|BC}+p_{2}\gamma_{B|AC}+p_{3}\gamma_{C|AB})\geq 0.
\end{equation}
Where $\gamma_{\cdot|\cdot\cdot}$ is block diagonal. We have denoted $\gamma_{A|BC}=\gamma_{A}\oplus\gamma_{BC}$ with $\gamma_{A}=diag(x,1/x)$ with $x>0$ specifying that the local state of $A$ mode is a squeezing vacuum state, and local state $BC$ is a two-mode squeezed state with squeezing parameter $s$. Hence
\begin{eqnarray}\label{sp10}
  \gamma_{A|BC}=\left(
                \begin{array}{ccc}
                x&0&0\\
                0&\cosh(2s) &\sinh(2s)\\
                0&\sinh(2s) &\cosh(2s) \\
                \end{array}
              \right)\nonumber\\
              \oplus\left(
                \begin{array}{ccc}
                \frac{1}{x} & 0 & 0 \\
                0 &  \cosh(2s) &-\sinh(2s)\\
                0 &  -\sinh(2s) &\cosh(2s) \\
                \end{array}
              \right)
\end{eqnarray}
Due to the symmetry of the three mode symmetric squeezed thermal state, we set $p_{1}=p_{2}=p_{3}=\frac{1}{3}$, $\gamma_{B|AC}$ and $\gamma_{C|AB}$ are transformations of $\gamma_{A|BC}$. Then criterion (\ref{sp9b0}) leads to
\begin{eqnarray}
   a-c\geq \frac{1}{3}[x+2\cosh(2s)-\sinh(2s)], \label{sp11}\\
   a+2c\geq \frac{1}{3}[x+2\cosh(2s)+2\sinh(2s)], \label{sp12} \\
   b+c\geq \frac{1}{3}[\frac{1}{x}+2\cosh(2s)+\sinh(2s)], \label{sp13}\\
   b-2c\geq\frac{1}{3}[\frac{1}{x}+2\cosh(2s)-2\sinh(2s)]. \label{sp14}
\end{eqnarray}
Utilizing $b=a+c$, then the left hand sides of (\ref{sp11}) and (\ref{sp14}) are equal. By equalizing the right hand side of (\ref{sp11}) and (\ref{sp14}) to obtain
\begin{equation}\label{sp15}
 x-\frac{1}{x}+\sinh(2s)=0.
\end{equation}
Then we have two independent inequalities (\ref{sp11}) and (\ref{sp12}) left. For larger $c$, (\ref{sp12}) is always true. We minimize the right hand side of (\ref{sp11}) with (\ref{sp15}). The solution is with $\sinh(2s)=\sqrt{\frac{9}{5+2\sqrt{11}}}$.  The biseparable condition is
\begin{equation}\label{sp16}
  a-c\geq \frac{4\sqrt{14+2\sqrt{11}}+\sqrt{29+8\sqrt{11}}-9}{6\sqrt{5+2\sqrt{11}}}.
\end{equation}
The right hand side of (\ref{sp16}) is approximately 0.812214. When $c$ becomes smaller, that is, $c<\frac{1}{\sqrt{5+2\sqrt{11}}}\approx 0.293190$, we need to consider both (\ref{sp11}) and (\ref{sp12}). The solution is $c=\frac{1}{3}\sinh(2s)$. So that the biseparable condition is
\begin{equation}\label{sp17}
  a\geq\frac{1}{2}\sqrt{c^2+\frac{4}{9}}+2\sqrt{c^2+\frac{1}{9}}-\frac{1}{2}c.
\end{equation}
The biseparable condition for the three mode symmetric squeezed thermal state with parameters $a$ and $c$ is the union of (\ref{sp16}) for $c\geq\frac{1}{\sqrt{5+2\sqrt{11}}}$ and (\ref{sp17}) for $c<\frac{1}{\sqrt{5+2\sqrt{11}}}$ as shown in Fig.\ref{Fig2}.

\section{Discussion and Conclusion}\label{sec7}

We used a Gaussian operator (detect operator), $\hat{M}$ with CM, $\gamma_{M}$ to detect the entanglement of either Gaussian states or non-Gaussian states. We provided three different methods to derive or refine separability criteria.

In the first method, we transformed the witness optimization to eigenvalue problem of an integral equation. The criterion obtained can be applied to multi-mode Gaussian states and some non-Gaussian states when they are cut into two parties under two additional conditions. One additional condition comes from that the characteristic function should fulfill the normalization condition of the subsystem state, $\chi_{A}(0)=tr(\rho_{A})=1$. This additional condition limits the freedom of detect operator choice, thus limits the entanglement detecting ability of the criterion obtained from integral equation of characteristic function. Another additional condition is that two submatrices derived from $\gamma_{M}$ should be diagonalizable by the same symplectic transformation, the condition comes from that if the integral equation of characteristic function can be decomposed to several basic two real variable integral equations. We proved that the eigenfunctions of a basic two real variable integral equation with symmetric Gaussian kernel is just the complete system of eigenfunctions of a harmonic oscillator, all the eigenvalues of the integral equation can be obtained so that the largest eigenvalue was obtained. It follows that the maximal mean of a Gaussian operator over product states is achieved by product Gaussian states. Thus the witness is a valid one. Applying the valid Gaussian witness to Gaussian state and non-Gaussian state prepared from Gaussian state by photon adding or subtracting leaded to necessary criteria of separability for these states, respectively. We provided two-mode symmetric Gaussian state and four mode ($2\times2$) symmetric Werner-Wolf state as examples. Further work can be done when the two submatrices $\delta$ and $\omega$ are not be simultaneously symplectic diagonalizable.

In order to lift the additional limitation on the witness of the first method, the second method is to numerically verify the hypothesis that the maximal mean of a Gaussian operator over product states is achieved by product Gaussian states. We dealt with the problem in Fock basis. We found that the problem can be simplified to verify if the product vacuum state can achieve the maximal mean of an operator over product states for the six parameter two mode Gaussian witness. We in fact proved that the two mode Gaussian witness with the six finite parameters is valid, although only numerically. It is independent of the analytic proof for infinite parameters in \cite{Chen22}. The framework we built here could be applied to multi-mode Gaussian state witness. We used two mode squeezed thermal state to illustrate the corresponding criterion of separability by the witness.

In the third method, we connect the separable criterion based on Gaussian witness to Werner-Wolf criterion. For Gaussian state entanglement detection, Werner-Wolf criterion is better than witness based criterion since our criterion in Proposition \ref{proposition5} is a result of Werner-Wolf criterion. However, our witness based criteria are easily be applied to non-Gaussian entanglement detection. We refined the Werner-Wolf criterion, so that explicit necessary and sufficient criteria expressed with elements of covariance matrix can be derived.

In summary, separable criteria based on Gaussian operator were given for Gaussian states and non-Gaussian states prepared from Gaussian states by photon adding or/and subtracting, respectively. In many cases, the criteria are necessary and sufficient for separability. Integral equation of characteristic function was given to prove the validity of Gaussian witness. Numeric calculation framework was introduced to illustrate that the maximal mean of Gaussian operator over product states is achieved by product Gaussian states, with the two mode Gaussian operator as an example. The witness based criterion was connected with Werner-Wolf criterion. We also refined Werner-Wolf criterion and provided application examples, which are explicit separable (or biseparable) criteria completely with state parameters for the two mode Gaussian states, $2\times2$ generalized Werner-Wolf bound entangled states, $1^{\times n}$ and $1\times2$ symmetric Gaussian states.

The method of characteristic function integral equation should not be limited to Gaussian case. Non-Gaussian kernel could be considered in future work.

\section*{Acknowledgement}
This work is supported by the National Natural Science Foundation  of China (Grant No.61871347)

\section*{ Appendix A: Mean of Gaussian detect operator over product states}
In this appendix, we consider the optimality of eigenvalue corresponding to Gaussian eigenfunction of the integral eigen-equation in the main text. The eigen-equation for characteristic function $\chi_{A}$ is (\ref{sp0}). For simplicity, we consider the case of simultaneously diagnoalizable of $\delta$ and $\omega$ by some symplectic transformation. Then $\gamma_{2M}$ of the two-fold Gaussian kernel can be decomposed as direct summation of $2\times2$ matrices, each of them represents a position correlation or a momentum correlation between the two parties. The two-fold Gaussian kernel is reduced to a product of basic Gaussian kernels.

A basic Gaussian kernel takes the form of $\kappa(x,y)=\exp[-\alpha(x^2+y^2)+2\alpha rxy]$. The integral equation
 \begin{equation}\label{sp0a}
\int\kappa(x,y)\varphi(y)dy=\mu\varphi(x)
\end{equation}
has an unnormalized eigenfunction $\varphi_{0}(x)=\exp(-\beta x^2)$ with $\beta=\alpha\sqrt{1-r^2}$. We will show that this Gaussian eigenfunction corresponds to the maximal eigenvalue $\mu_{0}=\sqrt{\frac{\pi}{\alpha+\beta}}$ of the integral equation. The test eigenfunction $\varphi_{1}(x)=x\exp(-\beta x^2)$  for the next eigenvalue $\mu_{1}$ can be written as $\varphi_{1}(x)=\frac{\partial}{\partial J}\exp(-\beta x^2+Jx)|_{J=0}$. So that $\int\kappa(x,y)\varphi_{1}(y)dy=\mu_{1}\varphi_{1}(x)$, with $\mu_{1}=\mu_{0}\frac{\alpha r}{\alpha+\beta}$. The eigenfunction $\varphi_{2}(x)$ for the eigenvalue $\mu_{2}=\mu_{0}(\frac{\alpha r}{\alpha+\beta})^2$ is a linear combination of  $x^2\exp(-\beta x^2)$ and $\varphi_{0}(x)$. The spectrum then is $\mu_{n}=\mu_{0}(\frac{\alpha r}{\alpha+\beta})^{n}$. We can check that $\Sigma_{n=0}^\infty \mu_{n}=\sqrt{\frac{\pi}{2\alpha(1-r)}}=\int\kappa(x,x)dx$. No doubt, $\mu_{0}$ is the largest eigenvalue. So that Gaussian eigenfunction gives the largest eigenvalue for the basic Gaussian kernel integral equation.

More strictly, it can be shown that the eigenfunction corresponding to the eigenvalue $\mu_{n}=\mu_{0}(\frac{\alpha r}{\alpha+\beta})^{n}$ is
\begin{equation}\label{sp0b}
  \varphi_{n}(x)=N_{n}e^{-\beta x^2}H_{n}(\sqrt{2\beta}x),
\end{equation}
where $N_{n}$ is the normalization and $H_{n}(\eta)=\frac{d^n}{ds^n}e^{-s^2+2\eta s}|_{s=0}$ is the Hermitian polynomial. Let $\xi=\sqrt{2\beta}y, \eta=\sqrt{2\beta}x$, then $\int\kappa(x,y)\varphi_{n}(y)dy=\frac{N_{n}}{\sqrt{2\beta}}\frac{d^n}{ds^n}\int e^{-\frac{1}{2\sqrt{1-r^2}}(\eta^2+\xi^2-2r\xi\eta)-\frac{1}{2}\xi^2-s^2+2\xi s}d\xi|_{s=0}$. After the integral, it can be put into the form of $\mu_{n}\varphi_{n}(x)$ with $\mu_{n}=\sqrt{\frac{\pi}{\alpha+\beta}}(\frac{\alpha r}{\alpha+\beta})^n$.   The integral equation (\ref{sp0a}) and the harmonic oscillator share the same complete system of eigenfunctions.

\section*{Appendix B: Criterion from upper bound of Gaussian detect operator}

In the above case of two-mode symmetric Gaussian state, the minimization $\min_{\hat{M}}\frac{\Lambda}{Tr(\hat{\rho}_G \hat{M})}$ is achieved at $M_{i}\rightarrow\infty$ while keeping $\det{(\gamma_M)}$ finite.

Consider the six parameter Gaussian detect operator $\hat{M}$ with CM in (\ref{we2a0}).
The characteristic function of $\hat{M}$ is $\chi_{M}(z)=\exp{(-\frac{1}{4}z\gamma_{M}z^{T})}$, with $z=(z_{1},z_{2},z_{3},z_{4})$. The eigenvalues of $\gamma_{M}$ are $\lambda_{1\pm}=\frac{1}{2}((M_{1}+M_{3}\pm\sqrt{(M_{1}+M_{3})^2-4(M_{1}M_{3}-M_{5}^2)}$ and similar expression for $\lambda_{2\pm}$ as a function of $M_{2},M_{4},M_{6}$. In the limit of infinite $M_{i}$ while keeping $M_{1}M_{3}-M_{5}^2$ finite, we have $\lambda_{1+}\approx M_{1}+M_{3}$ tending to infinite and $\lambda_{1-}\approx \frac{M_{1}M_{3}-M_{5}^2}{M_{1}+M_{3}}$ tending to $0$. Let $u_{j}$ $(j=1\pm, 2\pm)$ be the column eigenvectors of $\gamma_M$, define $v_{j}=zu_{j} $, we have $v_{1+}=\frac{1}{\sqrt{\lambda_{1+}}}(\sqrt{M_{1}}z_{1}+\sqrt{M_{3}}z_{3})$. The characteristic function $\chi_{M}(z)=\exp{(-\frac{1}{4}\sum_{j=1\pm,2\pm}\lambda_{j}v_{j}^2)}$ then is
\begin{equation}\label{sp2}
  \chi_{M}(z)=\frac{4\pi}{\sqrt{\lambda_{1+}\lambda_{2+}}}\delta(v_{1+})\delta(v_{2+}).
\end{equation}
Where Gaussian function $\exp{(-\frac{1}{4}\lambda_{1+}v_{1+}^2)}$ is approximated as $\sqrt{\frac{4\pi}{\lambda_{1+}}}\delta(v_{1+})$ (Here $\delta$ is the Dirac delta function) when $\lambda_{1+}\rightarrow\infty$, Gaussian function $\exp{(-\frac{1}{4}\lambda_{1-}v_{1-}^2)}$ tends to $1$ as $\lambda_{1-}\rightarrow 0$.

For any single mode pure states $|\psi_{A}\rangle$ and $|\psi_{B}\rangle$, we have
\begin{eqnarray*}
&&\int\chi_{A}(z_{1},z_{2})\overline{\chi}_{M}(z)\chi_{B}(z_{3},z_{4})[dz]=\frac{2}{\sqrt{M_{3}M_{4}}}\\
   &&\times\int\chi_{A}(z_{1},z_{2})\chi_{B}(-\sqrt{\frac{M_{1}}{M_{3}}}z_{1},-\sqrt{\frac{M_{2}}{M_{4}}}z_{2})\frac{dz_1dz_2}{2\pi}\\
   &&\leq\frac{2}{\sqrt{M_{3}M_{4}}}[\int|\chi_{A}(z_{1},z_{2})|^2\frac{dz_1dz_2}{2\pi}]^\frac{1}{2}\\
   &&\times[\int|\chi_{B}(-\sqrt{\frac{M_{1}}{M_{3}}}z_{1},-\sqrt{\frac{M_{2}}{M_{4}}}z_{2})|^2\frac{dz_1dz_2}{2\pi}]^\frac{1}{2}\\
   &&=\frac{2}{\sqrt[4]{M_{1}M_{2}M_{3}M_{4}}}=:\Lambda_{+}.
\end{eqnarray*}
Where we have used Cauchy-Schwartz inequality and $\int|\chi(z)|^2[dz]=1$ for a pure state. Notice that $\Lambda_{+}$ is an upper bound of $\Lambda$ since the equality in the Cauchy-Schwartz inequality may not be reached.

Let $\Xi$ be the set of $\gamma_{M}$ with $M_{i}\rightarrow\infty$ and finite $\det{(\gamma_M)}$. If $\gamma_{M}\in \Xi$, the Gaussian characteristic function $\chi_{M}(z)$ tends to a delta function. Then, Cauchy-Schwartz inequality can be applied to give an upper bound, $\Lambda_{+}$, for $\Lambda$
 regardless whether the pure product states are Gaussian or not \cite{Zhang}. It follows a necessary criterion of separability, which is given by
 \begin{equation}\label{we2d}
   \det(\gamma+\gamma_{M})\geq \frac{16}{\Lambda_{+}^2}=4\sqrt{M_{1}M_{2}M_{3}M_{4}},
 \end{equation}
for any given $\gamma_{M} \in \Xi$. For a two-mode Gaussian state with standard CM as described by (\ref{we2}) with $\mathcal{A}=diag(a,a), \mathcal{B}=diag(b,b), \mathcal{C}=diag(c_{1},-c_{2})$ and $c_{1}>0, c_{2}>0$, criterion (\ref{we2d}) leads to the following more explicit necessary criterion of separability \begin{equation}\label{we2d1}
  (\sqrt{ab}-c_{1})(\sqrt{ab}-c_{2})\geq 1.
\end{equation}

\end{document}